\documentclass[aps, pra, twocolumn, floatfix, nobalancelastpage, superscriptaddress, amsmath, amssymb, nofootinbib]{revtex4}

\usepackage{graphics}      
\usepackage{graphicx}      
\usepackage{longtable}     
\usepackage{bm}            
\usepackage{natbib}
\usepackage{physics}
\usepackage{xcolor}
\usepackage[caption=false]{subfig}
\usepackage{dcolumn}
\usepackage{hyperref}
\usepackage{comment}
\usepackage{pgf}

\makeatletter
\DeclareRobustCommand*{\bfseries}{%
   \not@math@alphabet\bfseries\mathbf
   \fontseries\bfdefault\selectfont
   \boldmath
}
\makeatother

\newcommand{\expectation}[1]{\left\langle #1 \right\rangle}

\newcommand{\balpha}{\bm{\alpha}}

\newcommand{\bmu}{\bm{\mu}}
\newcommand{\bfx}{\bm{x}}

\newcommand{\bfr}{\bm{r}}

\begin{document}

\title{Hadronic vacuum polarization correction to the bound-electron $g$~factor}

\author{Eugen Dizer}
\affiliation{Max Planck Institute for Nuclear Physics, Saupfercheckweg 1, 69117 Heidelberg, Germany}
\affiliation{Ruprecht Karl University of Heidelberg, Department of Physics and Astronomy, Im Neuenheimer Feld 226, 69120 Heidelberg, Germany}

\author{Zolt\'{a}n Harman}
\email[]{harman@mpi-hd.mpg.de}
\affiliation{Max Planck Institute for Nuclear Physics, Saupfercheckweg 1, 69117 Heidelberg, Germany}

\date{March 14, 2023}

\begin{abstract}

The hadronic vacuum polarization correction to the $g$~factor of a bound electron is investigated theoretically. An 
effective hadronic Uehling potential obtained from measured cross sections of $e^- e^+$ annihilation into hadrons 
is employed to calculate $g$ factor corrections for low-lying hydrogenic levels. Analytical Dirac-Coulomb wave functions,
as well as bound wave functions accounting for the finite nuclear radius are used. Closed formulas for the $g$ factor shift
in case of a point-like nucleus are derived. In heavy ions, such effects are found to be much larger than for the
free-electron $g$ factor.

\end{abstract}

\maketitle

\section{Introduction}
\label{sec:intro}

Precision Penning-trap experiments on the $g$~factor of hydrogenlike and few-electron highly charged ions allow a 
thorough testing of quantum electrodynamics (QED), a cornerstone of the standard model describing electromagnetic 
interactions. The $g$~factor of hydrogen-like silicon ($Z=14$) has been measured with a $5\times 10^{-10}$ relative 
uncertainty~\cite{Sturm2011, Sturm2013}, allowing to scrutinize bound-state QED theory (see e.g.~\cite{Pachucki2004, 
Pachucki2005, Karshenboim2002, Lee2005, Yerokhin2002, Yerokhin2004, Shabaev2002-2, Beier2000-1, Beier2000, Sailer2022,
Schneider2022}). Two-loop radiative effects and shifts due to nuclear structure and recoil are observable in such measurements. The high
accuracy which can be achieved on the experimental as well as theoretical side also enables the determination of
fundamental physical constants such as the electron mass $m_{\rm e}$~\cite{Sturm2014, Koehler2015, Zatorski2017,
Haeffner2000, Beier2002}. Recently, it was shown that $g$ factor studies can also help in the search for new physics,
i.e. the coupling strength of a hypothetical new interaction can be constrained through the comparison of theoretical and
experimental results~\cite{Debierre2020, Debierre2022, Sailer2022}.

Further improved tests and possible determinations of fundamental constants~\cite{Yerokhin2016, Shabaev2006, Cakir2020}
call for an increasing accuracy on the theoretical side. The evaluation of two-loop terms up to order $(Z\alpha)^{5}$
(with $Z$ being the atomic number and $\alpha$ the fine-structure constant) has been finalized recently~\cite{Czarnecki2018,
Pachucki2017}, increasing the theoretical accuracy especially in the low-$Z$ regime. First milestones 
have been also reached in the calculation of two-loop corrections in stronger Coulomb fields, i.e. for larger values of 
$Z\alpha$~\cite{Yerokhin2013-1, Sikora2020}. As the experiments are advancing towards heavy ions~\cite{Sturm2019,Kluge2008}, 
featuring smaller and smaller characteristic distance scales for the interaction between the bound electron and the nucleons, 
the effects of other forces may need to be considered as well.

Motivated by these prospects, in this article we investigate vacuum polarization (VP) corrections due to the virtual 
creation and annihilation of hadrons. The dominant VP contribution arises from virtual $e^- e^+$ pair creation, which 
has been widely investigated in the literature~\cite{Karshenboim2002, Lee2005, Karshenboim2001, Karshenboim2005} and 
is well understood. The other leptonic VP effect is due to virtual muons, the contribution of which is suppressed by 
the square of the electron-to-muon mass ratio~\cite{Landau1982}. The hadronic VP effect, which arises due to a superposition 
of different virtual hadronic states, is comparable in magnitude to muonic VP, however, it requires a completely different 
description since the virtual hadrons interact via the strong force. An effective approach to take into account such effects 
for the free-electron $g$~factor is described in e.g. Ref.~\cite{Burkhardt1989}, in which hadronic VP is characterized by the 
cross section of hadron production via $e^- e^+$ annihilation. Following this treatment, we apply the known empirical 
parametric hadronic polarization function for the photon propagator from Ref.~\cite{Burkhardt2001} to account for the 
complete hadronic contribution in case of the bound-electron $g$~factor.

\begin{figure}[h]
  \centering
  \subfloat[]{\includegraphics[width=0.16\textwidth]{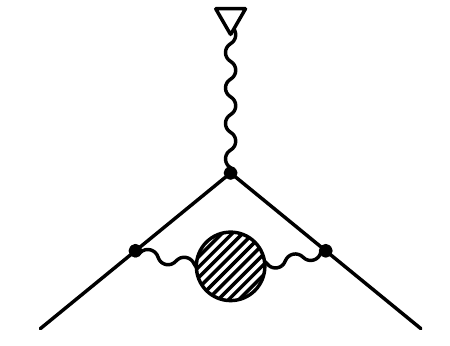}\label{fig:free_had_vp}}
  \subfloat[]{\includegraphics[width=0.32\textwidth]{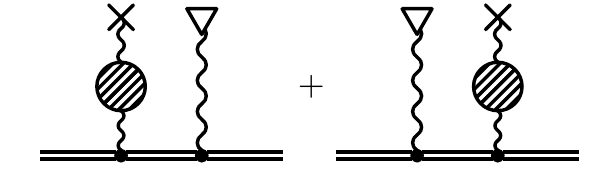}\label{fig:bound_had_vp}}
  \caption{Feynman diagrams representing the leading hadronic VP corrections to the free-electron $g$~factor 
  (\ref{fig:free_had_vp}) and the bound-electron $g$~factor (\ref{fig:bound_had_vp}). Double lines represent electrons 
  in the electric field of the nucleus and wavy lines with a triangle depict the interaction with the external magnetic field.
  For the free electron, it is a two-loop process where the self-interaction of the electron is perturbed by the effective 
  hadronic polarization function (shaded bubble). For the bound electron, it is a one-loop correction where the Coulomb
  interaction with the nucleus (cross) is perturbed by the effective hadronic polarization function.}
  \label{fig:had_vp_1}
\end{figure}

While in case of the free electron, the hadronic
correction only appears on the two-loop level, as a correction to the electrons  electromagnetic self-interaction 
(see Fig.~\ref{fig:free_had_vp}), in case of a bound electron it appears already as a one-loop effect (see Fig.~\ref{fig:bound_had_vp}). 
Furthermore, the hadronic VP is boosted by approximately $\sim Z^4$, i.e. by the fourth power of 
the nuclear charge number, and thus, as we will see later, for heavier ions above $Z=14$ its contribution is larger than in case of 
a free electron~\cite{Karshenboim2021}.

\clearpage

An effective potential constructed from the parametrized VP function, the hadronic Uehling potential, has been 
derived in Ref.~\cite{Breidenbach2022}. We calculate the perturbative correction to the $g$ factor due to this radial 
potential employing analytical Dirac-Coulomb wave functions, as well as numerically calculated wave functions 
accounting for a finite-size nucleus. Analytical formulas are presented, and numerical results are given for 
hydrogenic systems from H to U$^{91+}$.
We note that such an approach assumes an infinitely heavy nucleus, i.e. 
nuclear recoil effects are excluded in our treatment.

We use natural units with $\hbar=c=1$ for the reduced Planck constant $\hbar$ and the speed of light $c$, and 
$\alpha = {e^2}$, where $\alpha$ is the fine-structure constant and $e$ is the elementary charge. Three-vectors are 
denoted by bold letters.

\section{$g$ factor corrections}
\label{sec:theory}

Generally speaking, the $g$~factor describes the coupling of the electron's magnetic moment $\bmu$ to its total angular 
momentum $\bm{J}$. The corresponding first-order Zeeman splitting $\Delta{E}$ due to the electron's interaction with 
an external homogeneous magnetic field $\bm{B}$ is
\begin{equation}
  \label{eq:1}
  \Delta{E} = - \expectation{\bmu \cdot \bm{B}} = g \, \mu_{\mathrm{B}} \expectation{\bm{J} \cdot \bm{B}} \,,
\end{equation}
where $\mu_{\mathrm{B}} = e/(2m_{\rm e})$ is the Bohr magneton of the electron and $g$ is its $g$ factor, which depends on the 
electron configuration.

On the other hand, the relativistic interaction of an electron with the external magnetic field can be derived from the 
minimal coupling principle in the Dirac equation. In first-order perturbation theory, this leads to the energy shift
\begin{equation}
  \label{eq:2}
  \Delta{E} = e \expectation{\balpha \cdot \bm{A}} \,,
\end{equation}
where $\balpha$ are the usual Dirac matrices given in terms of the gamma matrices by $\alpha_i = \gamma^0 \gamma^i$~\cite{Peskin1995} 
and $\bm{A}$ is the vector potential for the magnetic field, such that $\bm{B} = \nabla \times \bm{A}$. Choosing the magnetic field to be 
directed along the $z$~axis, one can see that a possible choice for the vector potential is $\bm{A} = [\bm{B} \times \bfr]/2$, 
where $\bfr$ is the position vector. Together with Eq.~\eqref{eq:1} and \eqref{eq:2}, one can derive the following general 
expression for the $g$~factor~\cite{Karshenboim2001}:
\begin{equation}
  \label{eq:3}
  g = \frac{2\kappa m_{\rm e}}{j(j+1)} \int_0^{\infty} dr \, r G_{n\kappa}(r) F_{n\kappa}(r)  \,,
\end{equation}
where $n$ is the principal quantum number of the bound state, $j=\vert \kappa \vert - 1/2$ is the total angular momentum
quantum number and $\kappa$ is the relativistic angular momentum quantum number. The functions
$G_{n\kappa}(r), F_{n\kappa}(r)$ are the radial components in the electronic Dirac wave function,
\begin{equation}
  \label{eq:4}
  \psi_{n\kappa m}(\bfr) = \frac{1}{r} \left(
    \begin{array}{c}
      G_{n\kappa}(r) \ \Omega_{\kappa m}(\theta,\varphi)  \\
      i F_{n\kappa}(r) \ \Omega_{-\kappa m}(\theta,\varphi)  \\
    \end{array}
  \right)  \,,
\end{equation}
where $m$ is the magnetic quantum number and $r = \vert \bfr \vert$. The spherical spinors $\Omega_{\pm\kappa m}(\theta,\varphi)$
make up the angular components and are the same for any central potential $V(r)$~\cite{Johnson1988}.

A straightforward approach for calculating the $g$~factor shift $\Delta{g}^{\mathrm{VP}}$ due to vacuum polarization (VP), 
is to solve the radial Dirac equation numerically with the inclusion of the VP effect, and then substituting the perturbed 
functions $G_{n\kappa}^{\mathrm{VP}}(r), F_{n\kappa}^{\mathrm{VP}}(r)$ into Eq.~\eqref{eq:3}. The difference between the 
pertubed and the unperturbed $g$~factor gives the corresponding shift
\begin{equation}
  \label{eq:5}
  \Delta{g}^{\mathrm{VP}} = g^{\mathrm{VP}} - g  \,.
\end{equation}

However, we will apply a different method to investigate the hadronic $g$~factor shift. As shown in Ref.~\cite{Karshenboim2005}, 
owing to the properties of Dirac wave functions, the $g$~factor in Eq.~\eqref{eq:3} can be expressed through the energy eigenvalues $E_{n \kappa}$,
\begin{equation}
  \label{eq:6}
  g = - \frac{\kappa}{2j(j+1)} \left( 1 - 2\kappa \frac{\partial E_{n \kappa}}{\partial m_{\rm e}} \right)  \,,
\end{equation}
if the potential $V(r)$ does not depend on the electron mass $m_{\rm e}$. This formula was used successfully, e.g., to investigate 
the finite nuclear size effect in Ref.~\cite{Karshenboim2005}. We apply this new approach to investigate the vacuum 
polarization effect, described by an effective potential. Having a small perturbation $\delta{V}(r)$ to the nucleus potential 
(like the hadronic Uehling potential \cite{Breidenbach2022}), the $g$~factor shift can be shown to be~\cite{Karshenboim2005}
\begin{equation}
  \label{eq:7}
  \Delta{g}^{\mathrm{VP}} = - \frac{\kappa^2}{j(j+1)m_{\rm e}} \expectation{r \frac{\partial \delta V(r)}{\partial r}}  \,.
\end{equation}

For the relativistic ground state and a point-like nucleus, this expectation value can be evaluated further to obtain
\begin{equation}
  \label{eq:8}
  \Delta{g}_{1s}^{\mathrm{VP}} = \frac{4 (1+2\gamma)}{3m_{\rm e}} \Delta{E}_{1s}^{\mathrm{VP}} 
  - \frac{8 Z\alpha}{3} \expectation{r \delta V}_{1s}  \,,
\end{equation}
where $\gamma = \sqrt{1-(Z\alpha)^2}$ and $\Delta{E}_{1s} = \expectation{\delta V}_{1s}$ is the corresponding energy shift 
in first-order perturbation theory. Since the second term on the right-hand side of Eq.~\eqref{eq:8} is $Z\alpha$ times 
smaller than the first term, the $g$~factor shift can be approximated for light ions ($Z\alpha \ll 1$) with the formula:
\begin{equation}
  \label{eq:9}
  \Delta{g}_{1s}^{\mathrm{VP}} \approx \frac{4 (1+2\gamma)}{3m_{\rm e}} \Delta{E}_{1s}^{\mathrm{VP}}  \,.
\end{equation}
A similar expression also appeared in Ref.~\cite{Karshenboim2005,Cakir2020} in a different context, studying the finite size effect. 
However, we will investigate the applicability of this formula as an approximation for calculating the $g$~factor shift due 
to VP effects for light ions.

\subsection{Leptonic vacuum polarization correction to the $g$ factor}
\label{subsec:leptonic}

\begin{figure}
  \centering
  \subfloat[]{\includegraphics[width=0.15\textwidth]{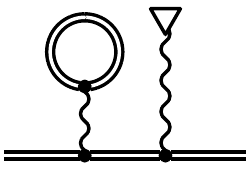}\label{fig:lept_vp_a}}
  \subfloat[]{\includegraphics[width=0.15\textwidth]{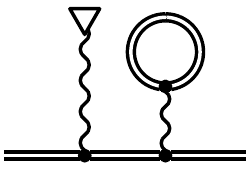}\label{fig:lept_vp_b}}
  \subfloat[]{\includegraphics[width=0.15\textwidth]{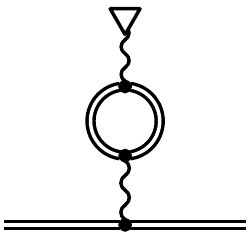}\label{fig:lept_vp_c}}
  \caption{Feynman diagrams representing the VP correction to the bound-electron $g$~factor. Double lines represent electrons 
  in the electric field of the nucleus and wavy lines with a triangle depict the interaction with the external magnetic field.}
  \label{fig:lept_vp}
\end{figure}

The leptonic VP correction to the bound-electron $g$ factor is well known. The corresponding diagrams are shown in 
Fig.~\ref{fig:lept_vp} and can be divided into two groups: the electric loop (EL) and the magnetic loop (ML) contribution. 
The vacuum polarization effect in the EL contribution (Fig.~\ref{fig:lept_vp_a} and Fig.~\ref{fig:lept_vp_b}) is equivalent 
to a perturbation in the interaction between the bound electron and the nucleus, and thus can be described by an effective 
perturbing potential $\delta{V}_{\mathrm{EL}}(\bfr)$. This allows the usage of perturbation theory and the simple inclusion 
of hadronic VP effects to the bound-electron $g$~factor shift, using Eq.~\eqref{eq:7}. 
As can be seen in Ref.~\cite{Lee2005, Belov2016}, the ML contribution (Fig.~\ref{fig:lept_vp_c}) is $Z\alpha$ times 
smaller than EL in the leading order, and is not the subject of the current work.

The vacuum loop in the EL contribution can be expanded in powers of the nuclear coupling strength $Z\alpha$, which corresponds 
to a free loop interacting with the nucleus. Due to Furry's theorem, only odd powers of $Z\alpha$ contribute~\cite{Furry1937, Belov2016}.
The leading term in this expansion is described by the Uehling potential $\delta{V}_{\mathrm{Ue}}(\bfr)$ and the contributions of higher
order in $Z\alpha$ are summarized to the Wichmann-Kroll potential $\delta{V}_{\mathrm{WK}}(\bfr)$, such that the effective perturbing 
potential is given by $\delta{V}_{\mathrm{EL}}(\bfr) = \delta{V}_{\mathrm{Ue}}(\bfr) + \delta{V}_{\mathrm{WK}}(\bfr)$~\cite{Belov2016}. 
The diagrams in Fig.~\ref{fig:lept_vp_a} and Fig.~\ref{fig:lept_vp_b} contribute equally to the EL correction. In this paper, 
we will investigate the leading contribution to the vacuum polarization due to the Uehling potential: 
$\delta{V}_{\mathrm{EL}}(\bfr) \approx \delta{V}_{\mathrm{Ue}}(\bfr)$.

In case of leptonic vacuum loops, the well-known leptonic Uehling potential is given by~\cite{Fullerton1976}
\begin{equation}
	\label{eq:10}
	\delta{V}_{\mathrm{Ue}}(\bfr) = - \frac{2\alpha (Z\alpha)}{3\pi} \int d^3x \, \rho(\bfx) \, 
	\frac{K_1(2 m_{\rm l} \vert \bfr - \bfx \vert)}{\vert \bfr - \bfx \vert} \,,
\end{equation}
where $\rho(\bfx)$ denotes the nuclear charge distribution normalized to unity, $m_{\rm l}$ is the mass of the virtual particle 
in the fermionic loop and $K_1(x)$ is given by
\begin{equation}
	\label{eq:11}
	K_1(x) = \int_1^{\infty} dt \ e^{-xt} \left( 1 + \frac{1}{2t^2} \right) \frac{\sqrt{t^2-1}}{t^2} \,.
\end{equation}

The $g$~factor shift of a bound electron in the ground state can be calculated analytically for a point-like nucleus and 
was already derived in~\cite{Karshenboim2001}. We will show that one arrives to the same result using the approach in 
Eq.~\eqref{eq:7}. Using the leptonic Uehling potential for a point-like nucleus ($\rho(\bfx) = \delta^{(3)}(\bfx)$)~\cite{Karshenboim2001},
\begin{equation}
	\label{eq:12}
	\delta{V}_{\mathrm{point}}^{\mathrm{lept.}}(r) = - \frac{2\alpha (Z\alpha)}{3\pi r} \ K_1(2 m_{\rm l} r) \,,
\end{equation}
and the radial components of the electronic wave function in the ground state~\cite{Landau1982}, one obtains from Eq.~\eqref{eq:7}
\begin{small}
\begin{align}
	\label{eq:13}
	\Delta{g}^{\mathrm{lept.}}_{\mathrm{point}}(1s) = - \frac{8 \alpha (Z\alpha)}{3 \pi s} \left[ I_{133} 
	- \frac{1}{3} I_{233} + \frac{Z \alpha s}{2\gamma} \left(I_{122} -  \frac{1}{3} I_{222} \right) \right] \,.
\end{align}
\end{small}
$I_{abc}$ is a modification of the base integral given in Ref.~\cite{Karshenboim2001}, see Appendix~\ref{sec:appendix_1}, 
and $s=m_{\rm e}/m_{\rm l}$ is the ratio of the electron and the loop particle masses.

The leading order $Z\alpha$ expansion is given by
\begin{align}
	\label{eq:14}
	\Delta{g}^{\mathrm{lept.}}_{\mathrm{point}}(1s) = \ &\frac{\alpha}{\pi} \left[- \frac{16 s^2 (Z\alpha)^4}{15} 
	+ \frac{5 \pi s^3 (Z\alpha)^5}{9} \right. \notag \\
	&+ \left(\frac{16s^2}{15}\ln(2sZ\alpha) -\frac{116s^2}{75} -\frac{16s^4}{7} \right) (Z\alpha)^6 \notag \\
	&+ \left(-\frac{5\pi s^3}{9}\ln\left(\frac{sZ\alpha}{2}\right) -\frac{8\pi s^3}{27} +\frac{7\pi s^5}{8} \right) (Z\alpha)^7 \notag \\
	&+ \mathcal{O}\left( (Z\alpha)^8 \right) \Big] \,.
\end{align}
For $s=1$, this is exactly the same result as in Ref.~\cite{Karshenboim2001}, however, obtained with a different method.
In the case of muonic VP, thus $s=m_{\rm e}/m_{\mu}$, the results for a finite size nucleus were obtained numerically in 
Ref.~\cite{Belov2016}.

In the next Subsection, we will use this approach to derive an analytic expression for the hadronic VP correction to the 
bound-electron $g$~factor.

\subsection{Hadronic vacuum polarization correction to the $g$ factor}
\label{subsec:hadronic}

As discussed in~\cite{Burkhardt1989, Burkhardt2001, Breidenbach2022}, the hadronic vacuum polarization function can be constructed semi-empirically 
from experimental data of $e^{-}e^{+}$ annihilation cross sections. The whole hadronic polarization function is parametrized 
for seven regions of momentum transfer and is given e.g. in Ref.~\cite{Burkhardt2001}. In Ref.~\cite{Breidenbach2022}, it was found that only the first region 
of parametrization is significant for the hadronic energy shift calculations. This is also clear from the physical point of 
view, since atomic physics is dominated by low energies around eV -- keV. Thus, we will use the analytic hadronic Uehling potential introduced 
in Ref.~\cite{Breidenbach2022} for our calculations. For a point-like nucleus it is given by
\begin{equation}
	\label{eq:15}
	\delta{V}^{\mathrm{had.}}_{\mathrm{point}}(r) = - \frac{2Z\alpha}{r} B_1 E_1\left( \frac{r}{\sqrt{C_1}} \right)  \,,
\end{equation}
with the coefficients $B_1 = 0.0023092$ and $C_1 = 3.9925370$ GeV$^{-2}$~\cite{Burkhardt2001, Breidenbach2022} 
and the exponential integral $E_1(x)$ which can be generalized for $n=0,1,2,...$ by~\cite{Abramowitz1972}
\begin{equation}
	\label{eq:16}
	E_n(x) = \int_1^{\infty} dt \ \frac{e^{-xt}}{t^n}  \,.
\end{equation}

The values for $B_1$ and $C_1$ are taken from the most recent parametrization in Ref.~\cite{Burkhardt2001} and will be used for the calculations.
The error of numerical results is estimated by comparison with an older parametrization in Ref.~\cite{Burkhardt1995} like has been
done in Ref.~\cite{Breidenbach2022}. \\

The corresponding hadronic Uehling potential for an extended nucleus with spherical charge distribution $\rho(\bfx)$ 
is obtained by the convolution~\cite{Breidenbach2022}
\begin{align}
	\label{eq:17}
	\delta{V}^{\mathrm{had.}}_{\mathrm{fns}}(\bfr) &= \int d^3x \ \rho(\bfx) \ \delta V^{\mathrm{had.}}_{\mathrm{point}}(\bfr-\bfx) \notag \\
	&= - \frac{4 \pi Z\alpha B_1 \sqrt{C_1}}{r} \int_0^{\infty} dx \ x \rho(x) D^{-}_2(r,x) ,
\end{align}
where $x = \vert \bfx \vert$ and
\begin{equation}
	\label{eq:18}
	D^{\pm}_n(r,x) = E_n\left( \frac{\vert r-x \vert}{\sqrt{C_1}} \right) \pm E_n\left( \frac{\vert r+x \vert}{\sqrt{C_1}} \right)  \,.
\end{equation}

As in our previous work~\cite{Breidenbach2022}, we will consider the homogeneously charged sphere as the model for the 
extended nucleus with root-mean-square (RMS) radii taken from Ref.~\cite{Angeli2013}. The charge distribution $\rho(r)$ is given by
\begin{align}
    \label{eq:19}
	\rho(r) = \frac{3}{4 \pi R^3} \, \theta(R-r)\,,
\end{align}
where $\theta(x)$ is the Heaviside step function and the effective radius $R$ is related to the RMS nuclear charge radius $R_{\mathrm{rms}}$ 
via $R = \sqrt{5/3} \, R_{\mathrm{rms}}$. The correspondig hadronic Uehling potential is given analytically 
in~\cite{Breidenbach2022}, see Appendix~\ref{sec:appendix_2}.

Let us turn to the evaluation of the leading hadronic VP contribution to the bound-electron $g$~factor, depicted in 
Fig.~\ref{fig:bound_had_vp}. In the low-energy limit, the hadronic Uehling potential is given by~\cite{Friar1999}
\begin{equation}
	\label{eq:20}
	\delta{V}^{\mathrm{had.}}_{\mathrm{non-rel.}}(\bfx) =  -4\pi Z\alpha B_1 C_1 \delta^{(3)}(\bfx) \,.
\end{equation}

Using Eq.~\eqref{eq:7} and the non-relativistic expectation value of the delta function, the leading order in $Z\alpha$ 
of the hadronic $g$~factor shift for general $ns$ states is found to be~\cite{Dizer2020}
\begin{align}
	\label{eq:21}
	\Delta{g}^{\mathrm{had.}}_{\mathrm{non-rel.}}(ns) 
	&= - \frac{4}{3m_{\rm e}} \expectation{12\pi Z\alpha B_1 C_1 \delta^{(3)}(\bfx)}_{ns} \notag \\
  	&= - \frac{16(Z\alpha)^4 m_{\rm e}^2}{n^3} B_1 C_1 \,.
\end{align}

For the $1s$ state, a fully relativistic expression for the point-like nucleus can be given. Using the hadronic Uehling 
potential in Eq.~\eqref{eq:15} and the relativistic wave function of the ground state, one obtains with Eq.~\eqref{eq:7}:
\begin{equation}
	\label{eq:22}
	\Delta{g}^{\mathrm{had.}}_{\mathrm{point}}(1s) = \frac{4}{3 m_{\rm e}} \Delta{E}^{\mathrm{had.}}_{\mathrm{point}}(1s) 
	- \frac{8 B_1 (Z \alpha)^2 (2 \lambda \sqrt{C_1})^{2 \gamma}}{3 \gamma (1+2\lambda \sqrt{C_1})^{2 \gamma}} \,,
\end{equation}
where $\lambda = Z \alpha m_{\rm e}$ and $\Delta{E}^{\mathrm{had.}}_{\mathrm{point}}(1s)$ is the analytical energy shift for a 
point-like nucleus given in Ref.~\cite{Breidenbach2022},
\begin{align}
	\label{eq:23}
	\Delta{E}^{\mathrm{had.}}_{\mathrm{point}}(1s) = &- \frac{Z\alpha \lambda (2 \lambda \sqrt{C_1})^{2 \gamma} B_1}{\gamma^2} \notag \\ 
	&\times {}_2F_1\left(2\gamma, 2\gamma; 1+2\gamma;-2 \lambda \sqrt{C_1}\right) \,,
\end{align}
with ${}_2F_1(a,b;c;z)$ being the hypergeometric function~\cite{Abramowitz1972}. 
The expansion of this expression up to $6^{\rm{th}}$ order in $Z\alpha$ is given by
\begin{align}
	\label{eq:24}
	\Delta{g}^{\mathrm{had.}}_{\mathrm{point}}(1s) = \ &- 16 B_1 C_1 m_{\rm e}^2 (Z \alpha)^4 
	+ \frac{512 B_1 C_1^{3/2} m_{\rm e}^3 (Z \alpha)^5}{9} \notag \\
	&- \frac{16 B_1 C_1 m_{\rm e}^2 (Z \alpha)^6}{3} \Big[ 2+30 C_1 m_{\rm e}^2 \notag \\
	&\left. \quad -3\ln(2 m_{\rm e} Z\alpha\sqrt{C_1}) \right] + \mathcal{O}\left( (Z\alpha)^7 \right) \,,
\end{align}
and it coincides with the non-relativistic approximation in Eq.~\eqref{eq:21} to order $(Z\alpha)^4$. \\

A similar relativistic calculation for the $2s$ state yields
\begin{align}
	\label{eq:25}
	\Delta{g}^{\mathrm{had.}}_{\mathrm{point}}(2s) = \ &- 2 B_1 C_1 m_{\rm e}^2 (Z \alpha)^4 
	+ \frac{64 B_1 C_1^{3/2} m_{\rm e}^3 (Z \alpha)^5}{9} \notag \\
	&- \frac{B_1 C_1 m_{\rm e}^2 (Z \alpha)^6}{24} \Big[ 41+420 C_1 m_{\rm e}^2 \notag \\
	&\left. \quad -48\ln(m_{\rm e} Z\alpha\sqrt{C_1}) \right] + \mathcal{O}\left( (Z\alpha)^7 \right) \,.
\end{align}

The leading orders of Eq.~\eqref{eq:24} and Eq.~\eqref{eq:25} satisfy the non-relativistic relationship in Eq.~\eqref{eq:21},
\begin{align}
	\label{eq:26}
	\Delta{g}^{\mathrm{had.}}_{\mathrm{non-rel.}}(ns) = \frac{1}{n^3} \Delta{g}^{\mathrm{had.}}_{\mathrm{non-rel.}}(1s) \,.
\end{align}

\subsection{Hadronic vacuum polarization correction to the reduced $g$ factor}
\label{subsec:reduced}

Additionally, we investigate hadronic effects on the weighted difference of the $g$ factor and the bound-electron energy $E$ 
of H-like ions, called reduced $g$ factor,
\begin{equation}
  \label{eq:27}
  \tilde{g} = g - \frac{4 (1+2\gamma)}{3m_{\rm e}} E  \,,
\end{equation}
put forward in Ref.~\cite{Cakir2020} for a possible novel determination of the fine-structure constant, and for testing
physics beyond the standard model~\cite{Debierre2022}. It was shown there that 
the detrimental nuclear structure contributions featuring large uncertainties can be effectively suppressed in
the above combination of the $g$ factor and level energy of the hydrogenic ground state. The question arises whether the 
same can be said about the hadronic VP corrections investigated in the present article.

The hadronic VP correction to the reduced $g$ factor for a point-like nucleus can be found analytically using Eq.~\eqref{eq:22}
and Eq.~\eqref{eq:23}. The leading order $Z\alpha$ expansion is given by
\begin{align}
	\label{eq:28}
	\tilde{g}^{\mathrm{had.}}_{\mathrm{point}}(1s) = \ &\Delta{g}^{\mathrm{had.}}_{\mathrm{point}}(1s) 
	- \frac{4(1+2\gamma)}{3 m_{\rm e}} \Delta{E}^{\mathrm{had.}}_{\mathrm{point}}(1s) \notag \\
	= \ &\frac{128}{9} B_1 C_1^{3/2} m_{\rm e}^3 (Z\alpha)^5 - 64 B_1 C_1^2 m_{\rm e}^4 (Z\alpha)^6 \notag \\
	&+ \mathcal{O}\left( (Z\alpha)^7 \right) \,.
\end{align}

Thus, the leading term of order $(Z\alpha)^4$ in $\Delta{g}^{\mathrm{had.}}_{\mathrm{point}}(1s)$ cancels such that the 
hadronic VP contribution to the reduced $g$ factor is indeed small for practical purposes. This also supports the approximation in Eq.~\eqref{eq:9}.
Therefore, we may conclude that hadronic effects do not hinder the extraction of $\alpha$ or detailed tests of QED and standard model extensions via
the measurement of $\tilde{g}$.

\subsection{Hadronic vacuum polarization correction to the weighted $g$ factor difference of H- and Li-like ions}
\label{subsec:weighted}

Another quantity of interest is the weighted difference of the $g$ factors of the Li-like and H-like charge states of the
same element,
\begin{align}
	\label{eq:29}
	\delta_{\Xi}g = g(2s) - \Xi \, g(1s) \,,
\end{align}
where $g(2s)$ is the $g$ factor of the Li-like ion and $g(1s)$ is the $g$ factor of the H-like ion. For light elements, 
the parameter $\Xi$ can be calculated to great accuracy by~\cite{Yerokhin2016,Yerokhin2016-1}
\begin{align}
	\label{eq:30}
	\Xi = 2^{-2\gamma-1} \left[1+\frac{3}{16}(Z\alpha)^2\right] \left(1-\frac{2851}{1000}\frac{1}{Z} 
	+ \frac{107}{100}\frac{1}{Z^2}\right) \,.
\end{align}
This weighted (or specific) difference was introduced to suppress uncertainties arising from the nuclear charge radius 
and further nuclear structural effects~\cite{Shabaev2002-1}. Therefore, bound-state QED theory can be investigated more accurately 
in $g$ factor experiments combining H- and Li-like ions than with the individual ions alone.

As we have seen, the hadronic VP correction to $\delta_{\Xi}g$ for a point-like nucleus can be found analytically. We approximate 
$\Delta{g}^{\mathrm{had.}}_{\mathrm{point}}(2s)$ of the Li-like ion with the expression in Eq.~\eqref{eq:25}
for the H-like ion. Since there are no electron-electron interactions in this approximation, we have to neglect
the terms of relative orders $1/Z$ and $1/Z^2$ in Eq.~\eqref{eq:30}. We note that the residual weight
\begin{align}
	\label{eq:31}
	\Xi_0 = 2^{-2\gamma-1} \left[1+\frac{3}{16}(Z\alpha)^2\right] \,,
\end{align}
exactly cancels the first two leading orders $(Z\alpha)^4$ and $(Z\alpha)^5$:
\begin{align}
	\label{eq:32}
	\delta_{\Xi_0}g^{\mathrm{had.}}_{\mathrm{point}} &= \Delta{g}^{\mathrm{had.}}_{\mathrm{point}}(2s) 
	- \Xi_0 \, \Delta{g}^{\mathrm{had.}}_{\mathrm{point}}(1s) \notag \\
	&= \frac{5}{2} B_1 C_1^2 m_{\rm e}^4 (Z\alpha)^6 + \mathcal{O}\left( (Z\alpha)^7 \right) \,.
\end{align}
Therefore, we can conclude that hadronic VP effects are also largely cancelled in the above specific difference. A similar conclusion can be
drawn for the case of the specific difference introduced for a combination
of H- and B-like ions~\cite{Shabaev2006}.
This result is well understood, since nuclear and hadronic VP contributions are both short-range effects with a similar behavior.

\section{Numerical Results}
\label{sec:result}

As mentioned in Ref.~\cite{Breidenbach2022}, the hadronic VP contribution to the energy shift is about $1/0.665 \approx 1.5$ times 
smaller than the muonic VP contribution in the case of the Uehling term. This can be also confirmed for the $g$ factor shift. 
Comparing the non-relativistic approximation for the hadronic $g$ factor shift $\Delta{g}^{\mathrm{had. \ VP}}_{\mathrm{non-rel., point}}$ 
in Eq.~\eqref{eq:20} with the first term of the expression for the muonic $g$ factor shift 
$\Delta{g}^{\mathrm{muonic \ VP}}_{\mathrm{non-rel., point}}$ in Eq.~\eqref{eq:14}, yields for hydrogen in the ground state
\begin{align}
	\label{eq:33}
	\Delta{g}^{\mathrm{had. \ VP}}_{\mathrm{non-rel., point}}(1s) &= -1.092(14) \times 10^{-16} \notag \\
	&= 0.664(9) \ \Delta{g}^{\mathrm{muonic \ VP}}_{\mathrm{non-rel., point}}(1s) \,.
\end{align}

\begin{figure}
  \centering
  \includegraphics[width=0.47\textwidth]{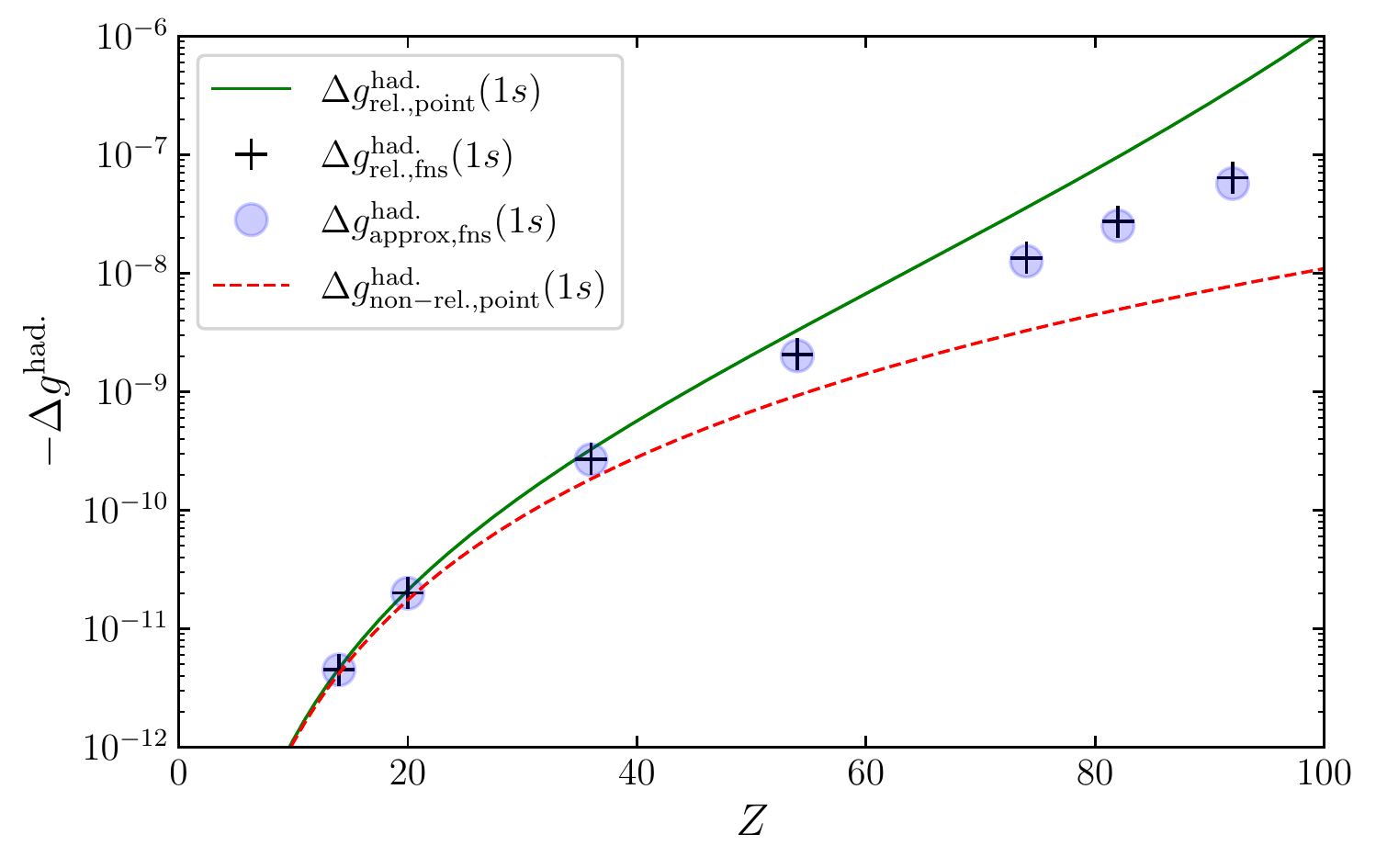}
  \caption{Comparison of analytical and numerical results for the hadronic $g$~factor shift of the bound electron in the ground state
  of H-like ions with atomic numbers $Z$ obtained in this work, see Table~\ref{tab:had_g_shift}. The green solid line represents the 
  analytical expression for a point-like nucleus $\Delta{g}_{\mathrm{rel., point}}^{\mathrm{had.}}$ in Eq.~\eqref{eq:21}, 
  while the red dashed line represents the non-relativistic expression $\Delta{g}_{\mathrm{non-rel., point}}^{\mathrm{had.}}$ 
  in Eq.~\eqref{eq:20}. The full numerical results for extended nuclei $\Delta{g}_{\mathrm{rel., fns}}^{\mathrm{had.}}$ 
  (crosses) are compared to the approximation $\Delta{g}_{\mathrm{approx, fns}}^{\mathrm{had.}}$ in Eq.~\eqref{eq:9} (circles) 
  with hadronic energy shifts for extended nuclei taken from Ref.~\cite{Breidenbach2022}.}
  \label{fig:hadronic}
\end{figure}

The values for the hadronic $g$ factor shift with an extended nucleus were calculated numerically using two different methods, 
both yielding the same results within the given uncertainties. The first method consists of calculating the expectation value
in Eq.~\eqref{eq:7} with the FNS hadronic Uehling potential and the semi-analytic wave functions of a homogeneously charged
spherical nucleus given in Ref.~\cite{Patoary2018}. As a consistency check, these results were reproduced by using the approach
of solving the radial Dirac equation numerically with the inclusion of the FNS potential, and substituting the resulting large
and small radial wave function components into Eq.~\eqref{eq:3} and Eq.~\eqref{eq:5}. The results for the hydrogenlike systems
H, Si, Ca, Xe, Kr, W, Pb, Cm and U are given in Table~\ref{tab:had_g_shift}. A diagrammatic representation is shown in
Fig.~\ref{fig:hadronic}. We note that for $Z=14$ and above, the magnitude of the hadronic vacuum polarization
terms considered in this work exceed in magnitude the hadronic contribution to the free-electron $g$ factor~\cite{Nomura2013}.
However, it is important to mention that the uncertainty of the leading finite nuclear size correction to the $g$ factor is approximately
an order of magnitude larger than the hadronic VP effect for all elements considered (see e.g.~\cite{Cakir2020}),
hindering the identification of the effect.

The errors given in Table~\ref{tab:had_g_shift} and~\ref{tab:had_g_shift_2s} are based on the uncertainty of the nuclear
root-mean-square radii $R_{\mathrm{rms}}$ given in Ref.~\cite{Angeli2013} and an assumed uncertainty for the parameters $B_1$ and $C_1$ 
as described in Section~\ref{subsec:hadronic}. The total error is dominated by the assumed uncertainty of $B_1$ and $C_1$. 
Owing to the closed analytical expression for the hadronic Uehling 
potential, numerical uncertainties are negligible. For the results $\Delta{g}_{\mathrm{approx, fns}}^{\mathrm{had.}}$ using the 
approximate formula in Eq.~\eqref{eq:9}, the hadronic energy shifts $\Delta{E}_{\mathrm{rel., fns}}^{\mathrm{approx}}$ from 
Ref.~\cite{Breidenbach2022} and their respective uncertainties are utilized.
For $Z=92$, the hadronic energy shift, which is not given in Ref.~\cite{Breidenbach2022}, was calculated using the same method.

One can see that the non-relativistic approximation in Eq.~\eqref{eq:21} represents a lower bound for the hadronic $g$ factor shift 
and is not sufficient for large atomic numbers $Z$. On the other hand, the analytic expression for the relativistic $g$ factor shift 
in case of a point-like nucleus in Eq.~\eqref{eq:22} represents an upper bound and differs also significantly from the numerical 
results for extended nuclei. We conclude that the effects due to a finite
size nucleus need to be included in a precision calculation of the hadronic
VP effect. At the present time, the uncertainty stemming from the assumed
nuclear charge distribution model limits the accuracy to 
about $1\%$~\cite{Breidenbach2022}. 
At the same time, the absence of more precise parametrizations of the hadronic polarization function in the low-energy regime 
limits the accuracy also to about $1\%$, see Table~\ref{tab:had_g_shift}. 
Thus, the given errors include, to a great part, all possible limitations of the uncertainty of the hadronic $g$ factor shift.

The simple approximate formula in Eq.~\eqref{eq:9} is found to be a good approximation for atomic numbers below $Z=14$. The error is 
less than $1\%$ for atomic numbers up to $Z=36$.

As shown in Section~\ref{subsec:reduced} and~\ref{subsec:weighted}, the hadronic VP contribution to the reduced and the
weighted $g$ factor in case of a point-like nucleus is at least $Z\alpha$ times smaller than the regular hadronic $g$ factor shift, see Eq.~\eqref{eq:24}.
In fact, numerical results for extended nuclei confirm that the hadronic contribution to both quantities does not differ significantly from
zero for small atomic numbers below $Z=36$ at the current level of accuracy. To see this, note that the numerical results for the finite-size
reduced and weighted $g$ factor can be obtained from Table~\ref{tab:had_g_shift} and~\ref{tab:had_g_shift_2s} via
\begin{align}
	\label{eq:34}
	\tilde{g}^{\mathrm{had.}}_{\mathrm{fns}}(1s) &= \Delta{g}_{\mathrm{rel., fns}}^{\mathrm{had.}}(1s)
	- \Delta{g}_{\mathrm{approx, fns}}^{\mathrm{had.}}(1s) \,, \\
	\delta_{\Xi_0}g^{\mathrm{had.}}_{\mathrm{fns}} &= \Delta{g}_{\mathrm{rel., fns}}^{\mathrm{had.}}(2s)
	- \Xi_0 \, \Delta{g}^{\mathrm{had.}}_{\mathrm{rel., fns}}(1s) \,,
\end{align}
respectively. For $Z=36$, one obtains
\begin{align}
	\label{eq:36}
	\tilde{g}^{\mathrm{had.}}_{\mathrm{fns}}(1s) &= -32(47) \times 10^{-13} \,, \\
	\delta_{\Xi_0}g^{\mathrm{had.}}_{\mathrm{fns}} &= -1(64) \times 10^{-14} \,.
\end{align}
Even for larger atomic numbers, hadronic effects do not constrain high-precision tests of QED
via the measurement of the reduced and weighted $g$ factor.

Recently, a high-precision measurement of the $g$~factor difference of two Ne isotopes was performed~\cite{Sailer2022}.
It was shown that QED effects mostly cancel, whereas nuclear effects like the nuclear recoil are well observable. In the following,
we investigate hadronic VP contributions to the bound-electron $g$ factor of the isotopes ${}^{20}$Ne$^{9+}$ and ${}^{22}$Ne$^{9+}$
in the ground state.

First, we calculate the hadronic VP correction to the $g$ factor difference stemming from the different nuclear size of the isotopes.
Nuclear recoil effects are excluded for now, and nuclear charge radii are taken from Ref.~\cite{Angeli2013}.
Using $R_{\text{rms}} = 3.0055(21)$ fm for ${}^{20}$Ne$^{9+}$ and $R_{\mathrm{rms}} = 2.9525(40)$ fm for ${}^{22}$Ne$^{9+}$,
the fully relativistic result for both isotopes is
\begin{align}
	\label{eq:38}
	\Delta g^{\mathrm{had.}}_{\mathrm{rel., fns}}\left(1s, {}^{20}\mathrm{Ne}^{9+}\right) &= - 1.133(14) \times 10^{-12} \,, \\
	\Delta g^{\mathrm{had.}}_{\mathrm{rel., fns}}\left(1s, {}^{22}\mathrm{Ne}^{9+}\right) &= - 1.133(15) \times 10^{-12} \,.
\end{align}
This is approximately a third of the hadronic contribution of the free electron given in Extended Table 1 in Ref.~\cite{Sailer2022}.
Thus, we conclude that at the given level of accuracy, hadronic effects of the bound electron also do not hinder the precise
calculation of the isotopic shift of ${}^{20}$Ne$^{9+}$ and ${}^{22}$Ne$^{9+}$.

To estimate also the hadronic VP correction stemming from the different nuclear mass of the isotopes
including nuclear recoil effects, we use the non-relativistic formula~\cite{Karshenboim2021}
\begin{align}
	\label{eq:40}
	\Delta g^{\mathrm{had.}}_{\mathrm{recoil}}(1s) &= \left( \frac{m_{\rm r}}{m_{\rm e}} \right)^2 \Delta g^{\mathrm{had.}}_{\mathrm{non-rel.}}(1s) \,,
\end{align}
with $m_{\rm r} = m_{\rm N} m_{\rm e} / (m_{\rm N} + m_{\rm e})$ being the reduced mass for an isotope with nuclear mass $m_{\rm N}$.
This is a reasonable approximation since the non-relativistic result for Ne ($Z=10$), using Eq.~\eqref{eq:21}, is
\begin{align}
	\label{eq:41}
	\Delta g^{\mathrm{had.}}_{\mathrm{non-rel.}}(1s, Z=10) &= - 1.092(14) \times 10^{-12} \,.
\end{align}
Using atomic masses from Ref.~\cite{Wang2012}, we obtain $m_{\rm r}({}^{20}\mathrm{Ne}^{9+}) = 0.99997 m_{\rm e}$ and
$m_{\rm r}({}^{22}\mathrm{Ne}^{9+}) = 0.99998 m_{\rm e}$, such that to first order:
\begin{align}
	\label{eq:42}
	\Delta g^{\mathrm{had.}}_{\mathrm{recoil}}(1s, {}^{20}\text{Ne}^{9+}) &= - 1.092(14) \times 10^{-12} \,, \\
	\Delta g^{\mathrm{had.}}_{\mathrm{recoil}}(1s, {}^{22}\text{Ne}^{9+}) &= - 1.092(14) \times 10^{-12} \,.
\end{align}
Thus, also the nuclear recoil effect to the hadronic VP contribution cannot be resolved at the given level of accuracy.

\begin{table*}
  \caption{Results for the hadronic VP contribution to the $g$~factor shift of the bound electron in the ground state 
  arising from the Uehling potential in the EL diagram (Fig.~\ref{fig:bound_had_vp}) using different approaches: the 
  non-relativistic approximation $\Delta{g}_{\mathrm{non-rel., point}}^{\mathrm{had.}}$ in Eq.~\eqref{eq:21}, 
  the relativistic formula for a point-like nucleus $\Delta{g}_{\mathrm{rel., point}}^{\mathrm{had.}}$ in 
  Eq.~\eqref{eq:22}, the approximate formula $\Delta{g}_{\mathrm{approx, fns}}^{\mathrm{had.}}$ using the hadronic 
  energy shift with an extended nucleus from~\cite{Breidenbach2022} in Eq.~\eqref{eq:9}, and the full relativistic
  result for an extended nucleus $\Delta{g}_{\mathrm{rel., fns}}^{\mathrm{had.}}$ using the analytical finite-size 
  Uehling potential with numerical finite-size wave functions in Eq.~\eqref{eq:7}.
  Root-mean-square nuclear charge radii $R_{\mathrm{rms}}$ are taken from~\cite{Angeli2013}.}
  \begin{ruledtabular}
    \begin{tabular}{llllll}
     $Z$ & $R_{\mathrm{rms}}$ [fm] & $\Delta{g}_{\mathrm{non-rel., point}}^{\mathrm{had.}}(1s)$ & $\Delta{g}_{\mathrm{rel., point}}^{\mathrm{had.}}(1s)$ & 
     $\Delta{g}_{\mathrm{approx, fns}}^{\mathrm{had.}}(1s)$ & $\Delta{g}_{\mathrm{rel., fns}}^{\mathrm{had.}}(1s)$
     \\
     \hline                                                                                                                       	 \\[-7pt]
     1  & 0.8783(86) & $-1.092(14) \times 10^{-16}$     & $-1.093(14) \times 10^{-16}$ & $-1.093(13) \times 10^{-16}$ & $-1.093(13) \times 10^{-16}$   \\
     14 & 3.1224(24) & $-4.196(53) \times 10^{-12}$     & $-4.616(57) \times 10^{-12}$ & $-4.490(56) \times 10^{-12}$ & $-4.497(56) \times 10^{-12}$   \\
     20 & 3.4776(19) & $-1.748(22) \times 10^{-11}$     & $-2.109(25) \times 10^{-11}$ & $-1.989(25) \times 10^{-11}$ & $-1.996(25) \times 10^{-11}$   \\
     36 & 4.1884(22) & $-1.835(23) \times 10^{-10}$     & $-3.263(39) \times 10^{-10}$ & $-2.664(33) \times 10^{-10}$ & $-2.696(34) \times 10^{-10}$   \\
     54 & 4.7859(48) & $-9.29(12)\,\,\,\times 10^{-10}$ & $-3.291(35) \times 10^{-9}$  & $-2.004(25) \times 10^{-9}$  & $-2.065(26) \times 10^{-9}$    \\
     74 & 5.3658(23) & $-3.275(41) \times 10^{-9}$      & $-3.568(32) \times 10^{-8}$  & $-1.261(15) \times 10^{-8}$  & $-1.344(17) \times 10^{-8}$    \\
     82 & 5.5012(13) & $-4.938(62) \times 10^{-9}$      & $-9.589(77) \times 10^{-8}$  & $-2.508(31) \times 10^{-8}$  & $-2.728(34) \times 10^{-8}$    \\
     92 & 5.8571(33) & $-7.825(98) \times 10^{-9}$      & $-3.572(24) \times 10^{-7}$  & $-5.705(71) \times 10^{-8}$  & $-6.410(80) \times 10^{-8}$    \\
    \end{tabular}
  \end{ruledtabular}
  \label{tab:had_g_shift}
\end{table*}

\begin{table}
  \caption{Results for the hadronic VP contribution to the $g$~factor shift of the bound electron in the $2s$ state.}
  \begin{ruledtabular}
    \begin{tabular}{lll}
     $Z$ & $\Delta{g}_{\mathrm{rel., fns}}^{\mathrm{had.}}(2s)$ & 
     \\
     \hline                                                                                                                       	 \\[-7pt]
     1  & $-1.366(17) \times 10^{-17}$  & \\
	 14 & $-5.673(71) \times 10^{-13}$  & \\
	 20 & $-2.542(32) \times 10^{-12}$  & \\
	 36 & $-3.583(45) \times 10^{-11}$  & \\
	 54 & $-2.966(37) \times 10^{-10}$  & \\
	 74 & $-2.189(27) \times 10^{-9}$   & \\
	 82 & $-4.728(59) \times 10^{-9}$   & \\
	 92 & $-1.213(15) \times 10^{-8}$   & \\
    \end{tabular}
  \end{ruledtabular}
  \label{tab:had_g_shift_2s}
\end{table}

\section{Summary}
\label{sec:summary}

Hadronic vacuum polarization corrections to the bound-electron $g$ factor have been calculated,
employing a hadronic polarization function constructed from empirical data on electron-positron
annihilation into hadrons. We have found that for a broad range of H-like ions, this one-loop effect is
considerably larger than hadronic VP for the free electron (see Fig.~\ref{fig:free_had_vp}).
Hadronic effects will be observable in future bound-electron $g$ factor experiments once nuclear charge
radii and charge distributions will be substantially better known. We have also found that the hadronic
effect does not pose a limitation on testing QED or physics beyond the standard model, and determining
fundamental constants through specific differences of $g$ factors for different ions, or through the
reduced $g$ factor. 
Finally, the analytic hadronic Uehling potential proves to be very useful and can be applied to further
atomic systems, e.g. positronium, or the hyperfine structure.

\section*{Acknowledgements}

E.~D. would like to thank the colleagues at the Max Planck Institute for Nuclear Physics,
especially the theory division lead by Christoph H. Keitel, for the hospitality during the work. 
We thank S. Breidenbach and H. Cakir for insightful conversations, and H. Cakir for assistance with numerical computations.
Supported by the Deutsche Forschungsgemeinschaft (DFG, German Research Foundation) – Project-ID 273811115 – SFB 1225.

\begin{widetext}
\appendix
\section{Base integral $I_{abc}$}
\label{sec:appendix_1}

The base integral $I_{abc}$ used in Eq.~\eqref{eq:13} is given by~\cite{Karshenboim2001}
\begin{align}
	\label{eq:base_int}
	I_{abc} = \ &\int_0^1 dy \ \frac{\left(1 - y^2\right)^{a - 1/2}}{y^{b-1}} \left( \frac{s Z \alpha y}{1 + s Z \alpha y}\right)^{c-2\epsilon} \notag \\
	= \ &\frac{1}{2} (s Z \alpha)^{c-2\epsilon} B\left(a+\frac{1}{2},1-\frac{b-c}{2}-\epsilon\right) \notag \\
	&\times {}_3F_2\left(\frac{c}{2}-\epsilon,\frac{c+1}{2}-\epsilon,1-\frac{b-c}{2}-\epsilon;\frac{1}{2},a+\frac{3-b+c}{2}-\epsilon;(s Z \alpha)^2\right) \notag \\
	&- \frac{c-2\epsilon}{2} (s Z \alpha)^{c+1-2\epsilon} B\left(a+\frac{1}{2},\frac{3-b+c}{2}-\epsilon\right) \notag \\
	&\times {}_3F_2\left(\frac{c}{2}+1-\epsilon,\frac{c+1}{2}-\epsilon,\frac{3-b+c}{2}-\epsilon;\frac{3}{2},a+2-\frac{b-c}{2}-\epsilon;(s Z \alpha)^2\right) \,,
\end{align}
where $s=m_{\rm e}/m_{\rm l}$ is the ratio of the electron and the loop particle masses, $\epsilon=1-\gamma$ with $\gamma = \sqrt{1-(Z\alpha)^2}$, $B(x,y)$ is the 
beta function and ${}_3F_2(a_1,a_2,a_3;b_1,b_2;z)$ is a generalized hypergeometric function~\cite{Abramowitz1972}.

\section{Hadronic Uehling potential for extended nuclei}
\label{sec:appendix_2}

The analytic hadronic Uehling potential for an extended nucleus with a spherical homogeneous charge distribution 
with effective radius $R$ is given by~\cite{Breidenbach2022} \\
\\ \
$r > R$:
\begin{align}
	\delta V^{\mathrm{had.}}_{\mathrm{fns, out}}(r)  = - \frac{3 Z \alpha B_1 \sqrt{C_1}}{r R^3}
	\left[  \sqrt{C_1} R \, D^{+}_3(r,R) - C_1 D^{-}_4(r,R) \right].
\end{align}
$r \leq R$:
\begin{align}
	\delta V^{\mathrm{had.}}_{\mathrm{fns, in}}(r) = - \frac{3 Z \alpha B_1 \sqrt{C_1}}{r R^3}
	&\left[ \sqrt{C_1} r + \sqrt{C_1} R E_3\left( \frac{r+R}{\sqrt{C_1}} \right)
	+ C_1 E_4\left( \frac{r+R}{\sqrt{C_1}} \right) \right. \notag \\
	&- \frac{1}{6} e^{\frac{r-R}{\sqrt{C_1}}} \left(2C_1 + \sqrt{C_1}(r+2R)+(r-R)(r+2R)\right) \notag \\
	&- \left. \frac{(r-R)^2(r+2R)}{6\sqrt{C_1}}E_1\left( \frac{R-r}{\sqrt{C_1}} \right)  \right].
\end{align}
The parameters $B_1$ and $C_1$ characterize the hadronic polarization function and are given in Section~\ref{subsec:hadronic}. \\
The functions $D^{\pm}_n(r,R)$ and $E_n(x)$ are defined in Eq.~\eqref{eq:18} and Eq.~\eqref{eq:16}, respectively. \\
\end{widetext}


\end{document}